# FibRace: a large-scale benchmark of client-side proving on mobile devices

**Simon Malatrait**
simon@kakarot.org
KKRT Labs

**Alex Sirac**
alex@hyli.org
Hyli

October 9, 2025

## Abstract

FibRace, jointly developed by KKRT Labs and Hyli, was the first large-scale experiment to test client-side proof generation on smartphones using Cairo M. Presented as a mobile game in which players proved Fibonacci numbers and climbed a leaderboard, FibRace served a dual purpose: to engage the public and to provide empirical benchmarking. Over a three-week campaign (September 11-30, 2025), 6,047 players across 99 countries generated 2,195,488 proofs on 1,420 unique device models. The results show that most modern smartphones can complete a proof in under 5 seconds, confirming that **mobile devices are now capable of producing zero-knowledge proofs reliably**, without the need for remote provers or specialized hardware. Performance was correlated primarily with RAM capacity and SoC (System on Chip) performance: devices with at least 3 GB of RAM proved stably, when Apple's A19 Pro and M-series chips achieved the fastest proving times. Hyli's blockchain natively verified every proof onchain without congestion. FibRace provides the most comprehensive dataset to date on mobile proving performance, establishing a practical baseline for future research in lightweight provers, proof-powered infrastructure, and privacy-preserving mobile applications.

# Contents

# Tables

# Figures

# 1 Introduction

Zero-knowledge proofs have become a cornerstone of blockchain scalability and privacy. In recent years, most progress in proving technology has focused on server-side proving, powering large-scale systems such as Layer 2 roll-ups and blockchain settlement layers. These developments, spurred by launches of high-throughput proving networks in 2024 and 2025, have dramatically reduced latency and cost for infrastructure-level proofs.

Yet this focus leaves a gap in a critical area: **client-side proving**.

Server-based models assume proofs are generated in centralized environments or distributed networks, but not on user devices. This architecture works well for throughput, not for privacy. Some use cases, especially those involving personal identity and private data, can only be realized if proofs are generated locally. Confidentiality is no longer guaranteed when sensitive data leaves the user's device for processing elsewhere.

Client-side proving minimizes friction for both developers and users. Developers no longer need to maintain or pay for remote-proving infrastructure: computation is delegated to end-user devices. Users, in turn, gain stronger privacy guarantees: proofs are generated and owned locally, without ever exposing private data to external servers or custodial intermediaries.

Recent incidents highlight this need vividly: as we write this report in October 2025, millions of Discord users' government IDs were leaked following a third-party verification breach, reigniting debates about how identity management should be handled online (Chia 2025). Similarly, systems like Google's zkWallet with Ligero and other emerging proof-of-age or proof-of-identity applications show that for many privacy-preserving use cases, the proof must be generated by the user, not by a remote prover (Kuhn 2025).

Despite the progress in ZK proving systems, client-side proving remains a technical challenge. For instance, a 2025 benchmark across several proving schemes showed that in-browser proving for p256 ECDSA signatures was still largely impractical, with only Noir achieving limited usability (Vlad 2025).

To achieve privacy and scalability at a global level, proof generation must become viable across the full spectrum of consumer devices, including low-end smartphones. KKRT Labs's **Cairo M**, a ZK-VM optimized for mobile hardware, was designed to address this challenge (KKRT Labs 2025).

Developing for client-side proving inevitably raises a central question: **Can we generate proofs directly on consumer devices, at scale, and under real-world conditions?**

In September 2025, **FibRace** was created as a joint initiative by **KKRT Labs**, developers of the Cairo M prover, and **Hyli**, the proof-powered blockchain for the next generation of applications. By disguising a client-side proving benchmark as a mobile game, it allowed us to test thousands of device models worldwide and begin answering this question experimentally.

# 2 Methodology

## 2.1 Overview

The FibRace campaign was designed as a **public mobile game** and a **controlled benchmark** of client-side proving using Cairo M.

Players interacted with a game interface, computing Fibonacci numbers and climbing leaderboards (Sirac 2025). Each interaction generated a data point for the study. This dual framing allowed the experiment to reach thousands of devices under natural, real-world usage conditions.

## 2.2 The FibRace gamified campaign

FibRace was a mobile game designed to turn a serious experiment in zero-knowledge benchmarking into an accessible and playful experience. Its core challenge was simple: prove Fibonacci numbers as quickly as possible and climb the leaderboard.

Each proof was generated locally on the player's smartphone using **Cairo M**, KKRT Labs's prover optimized for mobile hardware (Walter 2025). Once a proof was produced, it was transmitted to **Hyli's proof-powered blockchain**, which natively verifies Cairo M proofs and settles them onchain. Every successful proof advanced the player in the game and added a real data point to the benchmark.

The campaign ran from September 11 to 30, 2025, and reached a diverse global audience. Players from 99 countries participated across Android and iOS devices, 5,156 of them generating over 2 million proofs on 1,420 unique device models.

### 2.2.1 Gameplay

Participants could:

1. **Generate a proof** of the computation of the $n$-th term of the Fibonacci sequence, for a randomly chosen $n$.
2. **Mint a collector card** that displayed their number and proving time, designed for easy sharing on social media.
3. **Compete on leaderboards** tracking three metrics (fastest proof, highest proving frequency, and largest collection) to encourage repeated participation.

No financial rewards, airdrops, or tradeable assets were offered. This ensured that gameplay reflected genuine engagement and natural device diversity rather than speculative incentives.

### 2.2.2 Benchmark connection

Every player session generated measurable data collected automatically through the app's leaderboard and device-logging modules. This structure enabled KKRT Labs and Hyli to gather an unprecedented dataset on real-world proving performance.

By merging gaming dynamics with experimental rigor, FibRace shows that large-scale benchmarking of zero-knowledge provers can occur in open, decentralized conditions, engaging users not as test subjects, but as participants in a game.

## 2.3 Proof generation

Each game session consisted of the following sequence:

1. Sampling: The app pseudo-randomly sampled a $n$ in $[1..100,000]$ for which to compute the $n$-th term of the Fibonacci sequence $F_n$ on the M31 prime field.
2. Proof generation: The player's smartphone executed the Cairo M program locally to prove the correct computation of $F_n$, with 96 security bits. The higher the number $n$, the more resource-intensive the proving.
3. Verification and settlement: The proof was sent to Hyli, where it was natively verified and settled on the Hyli testnet blockchain.
4. Data logging: Metadata, including device model, OS version, RAM, SoC information, proof duration, and frequency, was recorded client-side and uploaded once verification succeeded.

All computation and proof generation occurred entirely on-device. No remote provers, cloud relays, or trusted execution environments were used.

## 2.4 Data collection and dataset structure

Four datasets were collected during the campaign, described in Table 1. The initial analysis was performed in a Jupyter Notebook, as detailed in (Malatrait 2025):

| Dataset | Description |
|---|---|
| mintedItems | One record per successfully generated and verified proof (2,195,488 rows) |
| players | Player profiles, including historical best proof and mint counters |
| devices | Hardware and OS metadata for each registered device (1,420 unique models) |
| crashLogs | Reports of failed or incomplete proof attempts, including device identifiers |

Table 1: Collected datasets during FibRace.

Each entry is timestamped.

## 2.5 Benchmark parameters and assumptions

The campaign defined a set of baseline assumptions to guide analysis and interpretation, summarized in Table 2.

A proof is deemed valid if successfully generated, sent to Hyli, verified by Hyli, and settled onchain on Hyli. The sampling window was the duration of the FibRace campaign, September 11 to 30, 2025. The program used for the benchmark would be a fixed Cairo M program computing the randomly chosen $n$-th term of the Fibonacci sequence.

| Parameter | Description | Assumed / Target Value |
|---|---|---|
| Minimum usable device RAM | We consider that client-side proving is viable if successful on enough devices with a RAM higher than this. | ≥ 3 GB |
| Acceptable crash rate | We considered Cairo M was successful on device models under a certain failure threshold. | < 1% failure rate |
| Target proof duration | We considered Cairo M offered a smooth user experience if the median proving time was under this threshold. | ≤ 5 s per proof |

Table 2: FibRace benchmarking assumptions.

KKRT Labs used these thresholds to evaluate Cairo M's performance. Hyli used them to characterize onchain verification throughput.

### 2.6 Ethical and experimental considerations

- **Voluntary participation:** Players opted in by installing the application. No rewards, airdrops, or financial incentives were offered, and this was made clear in the launch posts as well as in the application's interface.
- **Data minimization:** Only technical and performance metrics were collected. No personal identifiers were stored. FibRace did not enable account creation or data portability.
- **Cross-platform coverage:** Android and iOS were supported, ensuring SoC diversity: Apple Silicon, Qualcomm Snapdragon, MediaTek, Huawei Kirin, and Samsung Exynos.

### 2.7 Data availability

The performance report and raw datasets are publicly available at https://github.com/kkrt-labs/cairo-m/tree/main/docs/fibrace.

## 3 Benchmark results

### 3.1 Global participation

Across the 19-day campaign, **6,047 player names** were registered. While some players registered multiple player names (due to multi-device use or reinstalling the application), we will refer to them as « players » in the remainder of this report. Of these players, 5,156 generated at least one valid proof.

The players used **1,420 unique device models**. Device models are provided by the manufacturers themselves and read from the OS API. According to the app stores' data, the game was installed by players in **99 countries**. The Figure 1 is an heatmap of the number of FibRace downloads per country.

In total, **2,195,488 proofs** were generated on FibRace during the campaign.

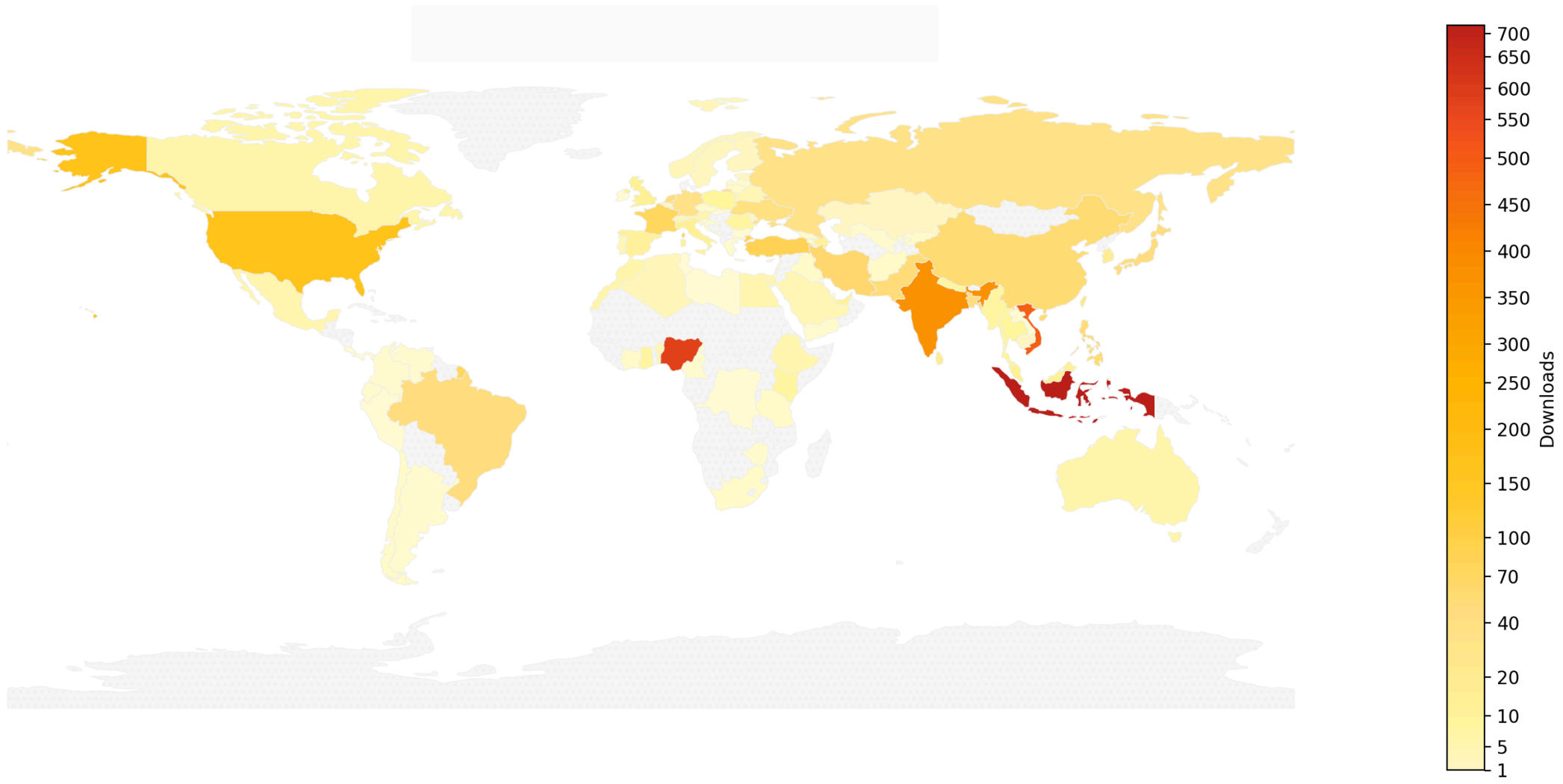

Figure 1: FibRace downloads by country.

## 3.2 Proving performance distribution

The duration of proof generation followed a right-skewed distribution, shown in Figure 2, with most devices completing a proof within **3 to 10 seconds**. The average frequency (Cairo M cycles per second) was 30 kHz, but the median was closer to 10 kHz, reflecting the heavy tail of slower devices.

High-performance clusters (≤ 3 s proofs) were led by the most recent iPhone and iPad models running Apple A19 Pro, M1–M4 chips, followed by older ones and high-end Android phones. Mid-range Android devices populated the 5–10 s band. Older devices (with an average proving time of more than 10s) showed increased crash incidence and throttling effects.

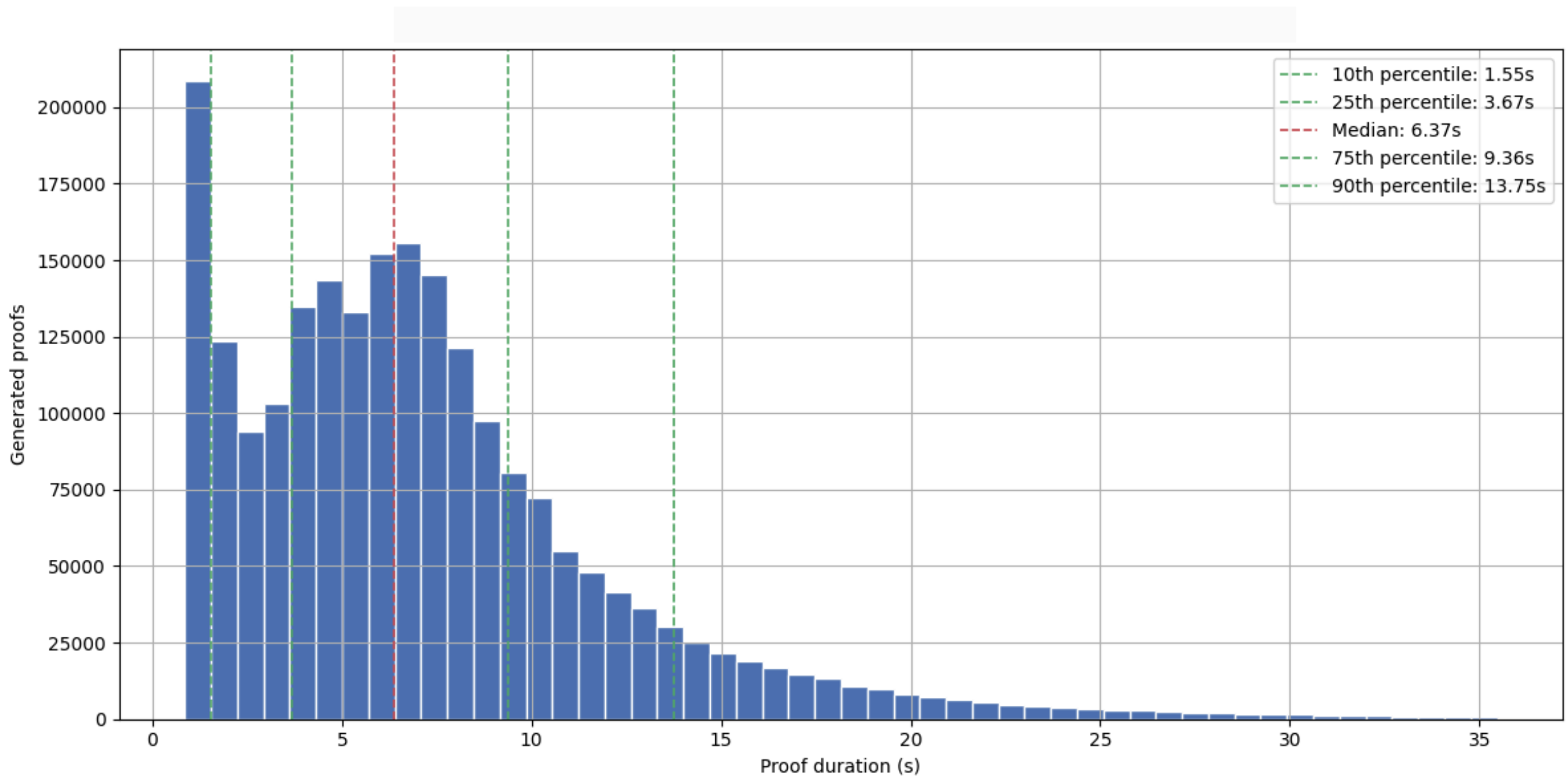

Figure 2: FibRace Cairo M proof generation duration distribution (central 99%).

## 3.3 Hardware correlations

### 3.3.1 RAM Capacity

Figure 3 displays key performance indicators, grouped by RAM buckets.

At least **3 GB** of available memory is required for successful proving on Cairo M.

Devices with **3–4 GB RAM** were the most prone to crashes due to Cairo M's memory footprint, while stability scaled steadily with RAM until plateauing beyond **6 GB**.

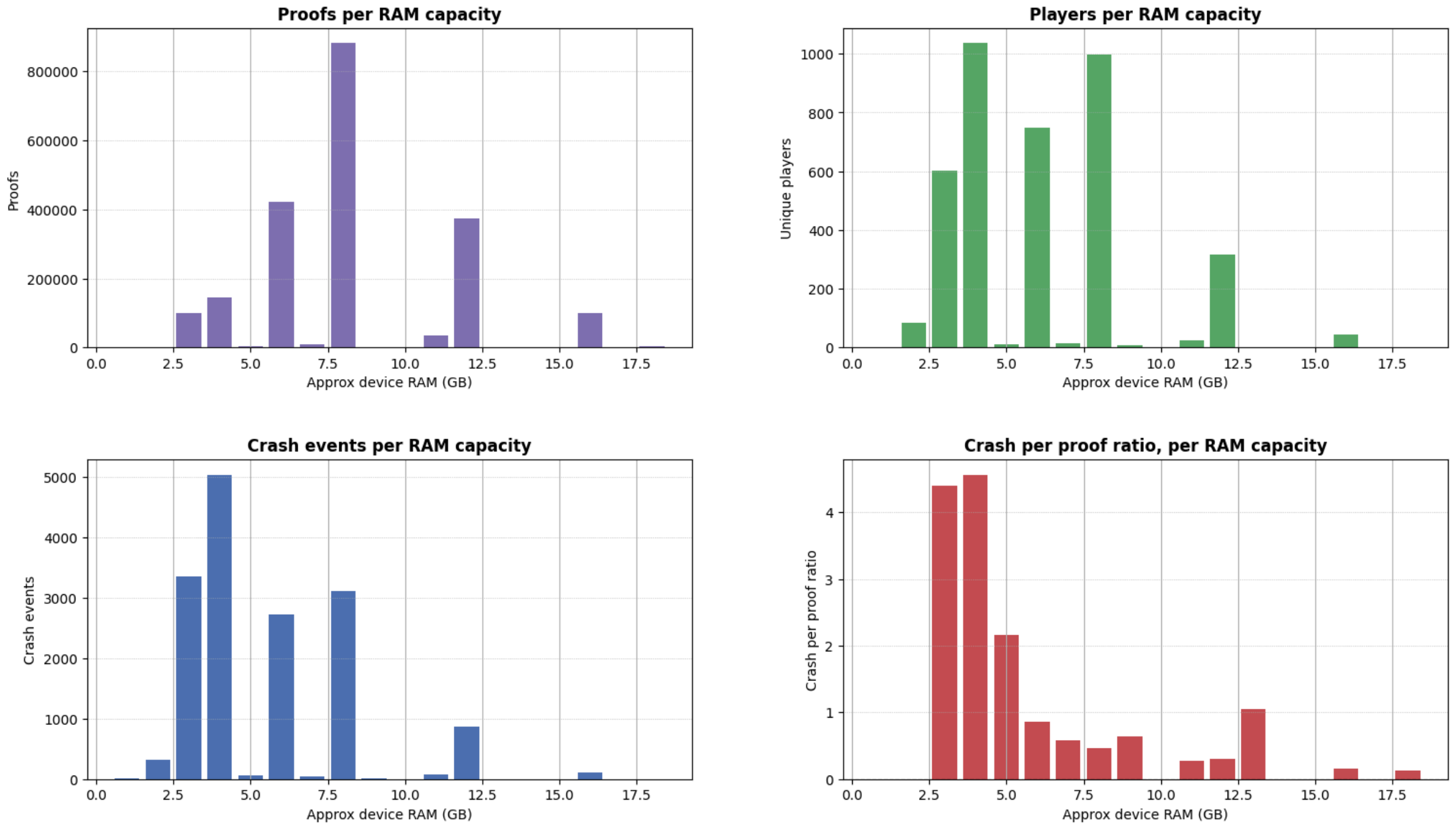

Figure 3: Key performance indicators per RAM capacity.

### 3.3.2 Device and manufacturer trends

Most proofs were generated on Xiaomi, Apple, and Samsung devices, as per Figure 4, reflecting global market share rather than platform bias (Kuzmin 2025). Within each brand, newer chip generations demonstrated predictable improvements in both duration and frequency.

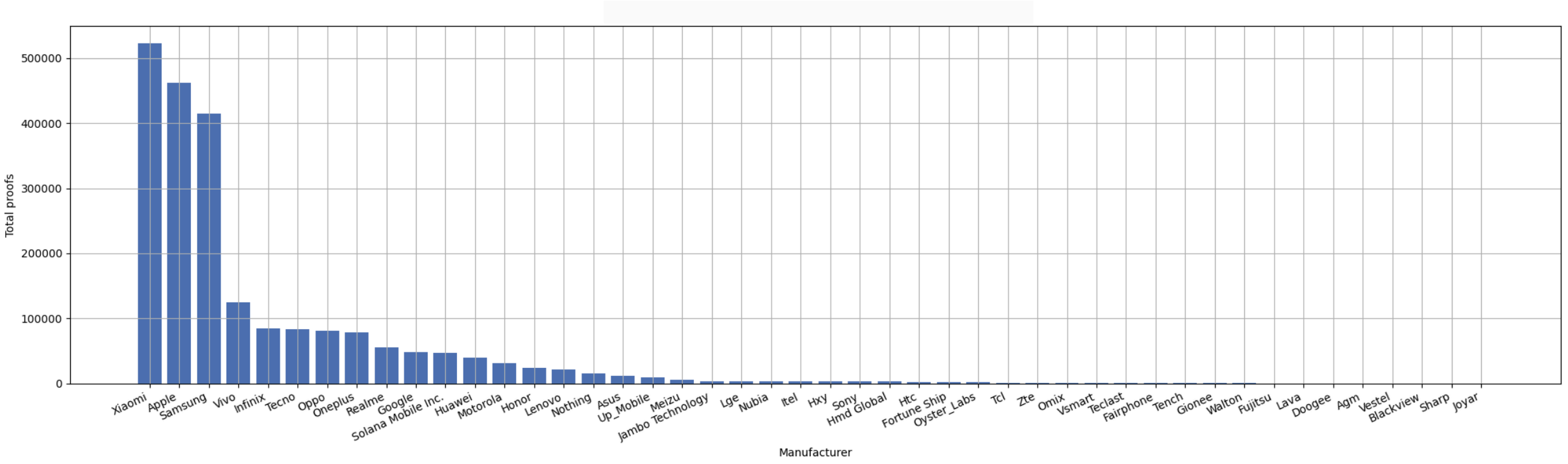

Figure 4: Proof volume per device manufacturer.

Figure 5 ranks the 50 best SoC, performance wise. It has been sorted based on proof duration as well as frequency, with best score as well as best average.

Apple SoCs dominated the upper performance tier, with the latest iPhone chips and M-series on iPad. iPads with the Apple M4 SoC and iPhones with the A19 Pro achieved the fastest proving times, approaching the limits of what Cairo M can deliver on mobile hardware.

Following the most recent Apple SoCs, high-end SoCs from Qualcomm, MediaTek, Kirin, and Exynos achieved great performance, with a proving time under 3 seconds.

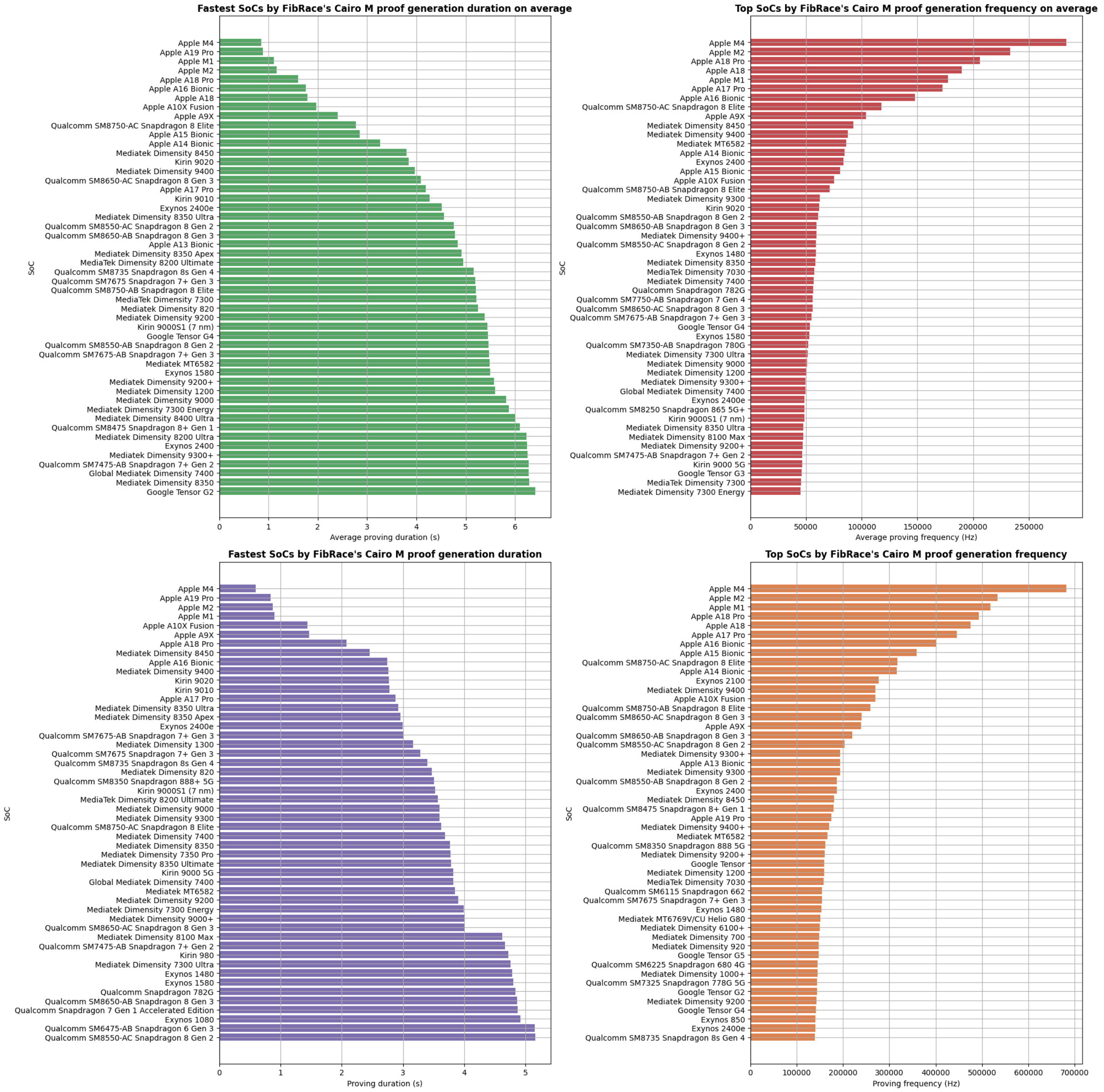

Figure 5: Best 50 SoCs for Cairo M proof generation.

## 3.4 Reliability and crashes

### 3.4.1 Reliability per device

Stable devices maintained success rates above 95%, confirming that **Cairo M can operate reliably on modern smartphones** without external proving support.

The overall crash rate during the FibRace campaign is **0.72%**, representing the number of recorded crash events (15,762 occurrences) over the total number of proof successfully generated. The majority of these occurred on devices with limited RAM (3-4 GB) as shown in Figure 3, which are often over five years old.

Crash rates correlated strongly with memory pressure rather than CPU limitations, suggesting that **future optimizations should prioritize reducing the RAM footprint**.

#### 3.4.2 Temporal trends

Hourly aggregates showed distinct activity peaks corresponding to regional evening hours, demonstrating genuine player engagement rather than automated testing.

Proof throughput and crash incidence followed inversely proportional patterns: as participation increased, minor rises in crash frequency indicated thermal throttling or sustained high-load conditions on some devices.

We also encountered a crash error on higher-RAM devices when minting many times in a row (usually more than a hundred). While unable to identify the root cause of these crashes with certainty, we assume that, although the spawned Rust thread ought to be killed at the end of each generated proof, it may not have been properly done, which could lead to OOM-type crashes. The short delay incurred when verifying and settling the proof onchain between two proof generation might be the cause. If this delay is too short, the thread may not properly close. Due to the short campaign duration and to the benchmark's focus on one-time proving, we did not investigate the issue further.

### 3.5 Summary interpretation

- **Performance:** Modern smartphones can generate Cairo M proofs in under 5 seconds on average.
- **Accessibility:** Proofs succeeded on 1,400+ device models, proving wide feasibility.
- **Reliability:** Devices with $\geq$ 3 GB RAM show consistent stability; below that threshold, proving fails systematically.
- **Hardware Insight:** CPU frequency correlates strongly with proof frequency, confirming Cairo M's effective scaling on mobile hardware.

## 4 Discussion and implications

### 4.1 Viability of client-side proving

The FibRace experiment demonstrates that client-side proving using **Cairo M** is technically viable and globally accessible.

Over 2 million proofs were generated, totaling 333 billions Cairo M cycles on more than 1,400 distinct device models, confirming that local proof generation can operate reliably across diverse hardware and network environments.

The results validate that the majority of modern smartphones can generate zero-knowledge proofs in under 5 seconds, which is well within the user-acceptable latency for casual or transactional applications. Devices with less than 3 GB of RAM consistently failed to complete a proof, delineating a clear lower bound for current implementations.

From a research perspective, FibRace provides the first large-scale, open dataset that quantifies the distribution of proof generation times across consumer hardware. This enables empirical calibration of future prover optimizations and helps refine benchmarks for memory footprint, cycle counts, and acceptable crash rates.

## 4.2 Implications for developers and ecosystem builders

For developers, these results are a strong signal that end-user devices can handle proving workloads previously reserved for servers or specialized hardware.

Client-side proving simplifies architecture, reduces cost, and improves privacy, aligning with the long-term goal of **self-sovereign computation**.

In practical terms, this means applications can now:

- Deploy zero-knowledge solutions without relying on remote provers.
- Execute verifiable interactions offline, syncing proofs asynchronously to Hyli.
- Build new interaction models (e.g., games, attestations, on-device identity proofs) that no longer depend on cloud resources.

For ecosystem builders, FibRace establishes a methodology for gamified benchmarking, collecting real-world performance data in a playful and opt-in format. This approach can be replicated for other proving schemes or for measuring device-specific bottlenecks, such as GPU acceleration or energy efficiency.

## 4.3 Broader impact

FibRace bridges the gap between cryptographic research and user experience. Re-framing benchmarking as a public game showcased how ZK technology can reach mainstream audiences without compromising scientific rigor.

Future iterations of this model could benchmark additional proving schemes, introduce controlled parameters (e.g., energy consumption, thermal throttling), and work upon different circuits.

## 4.4 Implications for Hyli and KKRT Labs

For Hyli, the experiment validated the scalability of its onchain proof verification layer. Over two million verifications were processed without congestion, even during peaks with 260,000 proofs per hour, demonstrating that Hyli's implementation of the Autobahn consensus can absorb high-frequency proof submissions from a large user base (Giridharan et al. 2024).

For KKRT Labs, FibRace served as a proof-of-concept for Cairo M, showing that its mobile optimization can sustain multi-million-proof throughput across everyday devices. The benchmark offers valuable feedback for future improvements.

Together, Hyli and KKRT Labs demonstrated that a proof-powered L1 and a mobile-optimized prover can function seamlessly, paving the way for scalable, privacy-preserving applications that can be deployed to consumer hardware.

# 5 Limitations of the experiment

While FibRace demonstrates the viability of client-side proving on mobile devices, several limitations must be acknowledged to correctly interpret the dataset and results.

## 5.1 Technical limitations

### 5.1.1 Benchmark circuit scope

The experiment focused exclusively on proving computations of terms of the Fibonacci sequence, an arithmetic circuit with low memory usage.

While representative for testing Cairo M baseline performance, and while we benched a wide range of witnesses, resulting in traces of 8 to 800,000 Cairo M cycles, it does not capture all of the Cairo M capabilities that can be found in real-world proving workloads, such as SHA2 hashes.

### 5.1.2 Memory footprint

Cairo M's memory consumption constrained participation on low-end devices. Proofs could not be generated on devices with less than 3 GB of RAM, and devices with 3–4 GB of RAM showed the highest crash rate, respectively 4.2 and 4.4 crashes per generated proof, due to OOM errors on the largest witnesses ($n$ near the 100,000 upper-bound).

It showcases that devices with limited available memory can generate proofs for most workloads of the benchmark, but not for the most expensive ones, due to memory constraints.

### 5.1.3 Energy and thermal effects

FibRace did not record power draw or thermal throttling. Anecdotal player feedback suggested noticeable heat and battery drain during extended play sessions, but these factors were not quantified.

Other performance factors were not measured, including battery state, energy saving mode activation, and CPU temperature.

Therefore we do not have a quantitative performance degradation ratio on our assumption that these factors lead to degraded performance.

Future benchmarks should include energy-use metrics to more effectively evaluate sustainability, user comfort and performance impact.

## 5.2 Game design limitations

### 5.2.1 Gameplay constraints

Because FibRace was designed as a game, participants could interrupt or close the app at any point, including by force-quitting the app while generating or sending a proof. Such interruptions are indistinguishable from technical crashes in the dataset, introducing ambiguity in interpreting some failure events.

### 5.2.2 No user accounts

The game operated without persistent user accounts to reduce friction and prevent airdrop speculation. While this ensured privacy and simplicity, it made longitudinal tracking of player behavior and device switching impossible. This did not affect the benchmark, but made it less enjoyable for some players.

### 5.3 Summary

These limitations do not diminish the core finding that modern smartphones can locally perform zero-knowledge proving. They do, however, highlight key areas for future experiments:

- measuring more complex circuits,
- measuring circuits with real-life applications,
- quantifying energy and heat effect on performance.

Together, these refinements would strengthen the empirical foundation laid by FibRace and support the next generation of client-side proving research.

## 6 Conclusion

FibRace demonstrates that client-side proving is no longer theoretical: it is operational, measurable, and already viable across a wide range of consumer devices.

Over the course of three weeks, more than 5,000 participants generated 2.2 million Cairo M proofs on 1,420 unique device models, confirming that proof generation can occur reliably on modern smartphones without the need for centralized infrastructure or trusted hardware.

The results validate a key assumption behind both KKRT Labs and Hyli: that distributed proving and native onchain verification can coexist efficiently. Cairo M made on-device proof generation accessible even to mid-range devices, while Hyli's proof-powered Layer 1 absorbed millions of verifications without network congestion.

Beyond its immediate findings, FibRace established a new model for open benchmarking through gamified participation. By blending experimentation and gameplay, it produced one of the most extensive public datasets on client-side proving to date, an approach that future ZK ecosystems can replicate for other circuits, provers, or device classes.

The experiment's limitations define clear next steps. Expanding to real-world use cases, and integrating thermal and battery instrumentation.

Ultimately, FibRace offers a simple but powerful insight: **zero-knowledge proofs can now run on users' own devices**.

This shift opens a new frontier for privacy-preserving applications and proof-native blockchains, where the path to scalable cryptography becomes user-driven.